# Transport properties around the metal-insulator transition for SrVO$_3$ ultrathin films fabricated by electrochemical etching

Hikaru Okuma, Yumiko Katayama, Keisuke Otomo, and Kazunori Ueno
Department of Basic Science, University of Tokyo
Meguro, Tokyo 153-8902, Japan

## Abstract

By using electrochemical etching, we fabricated conductive ultrathin SrVO$_3$ (SVO) films that exhibited metallic behavior down to 3 monolayers (ML). From an observed systematic change in transport properties with decreasing film thickness, it was found that the disorder in the films remained nearly unchanged during etching, and only the thickness was reduced. This is in contrast to the insulating behavior found for as-deposited SVO ultrathin films. For the etched films, the electron mobility at 200 K decreased with decreasing film thickness below 10 ML, originating from an increased scattering rate and electron effective mass near the metal-insulator transition. A slight upturn in the resistivity and a positive magnetoresistance at low temperatures were typically observed for the etched films down to 3 ML, which was explained by weak anti-localization of electrons in a weakly disordered metal.

## I. INTRODUCTION

A strongly correlated electron system shows a Mott transition due to competition between on-site Coulomb repulsion ($U$) and a one-electron bandwidth ($W$) [1,2]. The system becomes a Mott insulator when $U/W$ exceeds a critical value. In addition, strongly correlated metals in the vicinity of an insulator undergo not only a Mott transition but also Anderson localization due to crystal disorder [3]. Therefore, it is difficult to determine whether electron-electron interactions or disorder is the dominant driving force behind the metal-insulator transition (MIT).

SrVO$_3$ (SVO) ($a$ = 0.3843 nm) has a 3d$^1$ electronic configuration for vanadium and is, thus, a good candidate for studying the MIT in the metallic regime. The control parameters for the Mott transition are bandwidth and band filling [4]. In earlier studies, the MIT was mainly tuned by chemical substitution. For example, the band-filling controlled MIT in the La$_{1-x}$Sr$_x$VO$_3$ system is induced by an aliovalent A-site substitution [5-7]. On the other hand, the bandwidth change in the Ca$_{1-x}$Sr$_x$VO$_3$ (CSVO) system has been studied by isovalent A-site substitution with different ionic radii [8-11]. Taking into account $W$, CaVO$_3$ (CVO) is expected to be the most insulating. However, CSVO with $x$ = 0.2-0.5 is closest to the insulator in the whole composition range [10,11]. This indicates that the dominant driving force for the insulating state is the disorder induced by the solid solution between CVO and SVO. Then, it is necessary to study the electron-electron interaction effect on the MIT to change $W$ with fixed disorder.

In recent years, major improvements in molecular beam epitaxy in oxides have provided high-quality ultrathin films for exhibiting novel physical phenomena that uniquely emerge at heterointerfaces [12,13]. Furthermore, precise thickness control makes it possible to control $W$ by reducing the electronic dimensionality from 3 dimensions (3D) to 2 dimensions (2D). Then, we can isolate a change in $W$ from other competitive effects induced by chemical substitution [14-22]. However, studies using a few nanometer thick films showed that the critical thicknesses for the MIT differed by the degree of disorder, such as tensile strain at an interface [14] and surface degradation [18]. In addition, an abrupt increase in resistivity ($\rho$) at low temperature has also been reported around the critical thicknesses due to disorder, such as oxygen deficiency [15], surface scattering [20], and lattice strain near the magnetic transition [22]. Therefore, such disorder must be controlled for studying the MIT with a change in $W$ in ultrathin films.

For SVO thin films, *in situ* photoemission (PES) showed that Mott insulating states are induced with a thickness of 2-3 monolayers (ML) [23]. Transport studies also showed insulating behavior with an abrupt increase in resistivity in SVO films. However, the critical thickness for the insulator varies between 3 and 8 ML, all of which are thicker than 2 ML [17,21,24]. This suggests that the disorder-induced Anderson transition occurs at different thicknesses because each film has different disorder.

In this study, we eliminated differences in disorder between samples by using the electrochemical etching method [25-28]. Disorder in a thin film sample remained unchanged during electrochemical etching, but the dimensionality was changed by a reduction in the film thickness. In addition, the electrochemically etched ultrathin film shows no surface degradation because the surface remains coated with an ionic liquid (IL). We applied an electric double layer transistor (EDLT) configuration to SVO films [27,28]. After fixing the disorder, electronic transport properties as a function of the film thickness near the MIT were examined. The etched SVO films with a thickness of 3 ML exhibited conductive behavior, whereas the as-deposited films became good insulators. Then, we elucidated the transport properties of the ultrathin conductive films in the vicinity of the Mott insulator states by measuring the temperature dependence of $\rho$ and magnetoresistance (MR) at low temperature.

## II. EXPERIMENT

SVO thin films were deposited on (100)-oriented $(LaAlO_3)_{0.3}(Sr_2AlTaO_6)_{0.7}$ (LSAT) ($a$ = 0.3868 nm) and $SrTiO_3$ (STO) ($a$ = 0.3905 nm) substrates by the pulsed laser deposition method at $10^{-7}$ Torr and 900 °C. A ceramic target composed of SVO was ablated by a KrF excimer laser ($\lambda$ = 248 nm) with a repetition rate of 5 Hz and an energy fluence of 0.97-1.06 J/cm$^2$. Epitaxial growth of SVO (100) thin films was confirmed by an out-of-plane $2\theta/\omega$ scan and reciprocal space mapping with X-ray diffraction and XRD (Rigaku, Smart Lab). The thicknesses of the as-deposited films were determined by the deposition rate and confirmed with X-ray Laue oscillations.

We fabricated EDLTs with a thickness of 35 ML on as-deposited films (S1～S3) on LSAT. The SVO films were patterned into Hall bars with Au/Ti electrodes, and a Pt wire was placed over the channel and used as a gate electrode. The channel and Pt gate electrodes were covered with an IL droplet, N,N-diethyl-N-methyl-N-(2-methoxyethyl)ammonium bis(trifluoromethanesulfonyl)imide (DEME-TFSI).

The SVO thin films were repeatedly etched at 230 K with a gate bias ranging between -3 and -4.5 V, and the transport properties for the respective thicknesses ($t_{etch}$) were obtained. We also obtained transport properties for the as-deposited films with various thicknesses ($t_{depo}$). The transport properties were examined by using the four-probe method and a Hall measurement in a Physical Properties Measurement System (Quantum Design, PPMS) from 2 to 300 K and –3 to 3 T.

## III. RESULTS AND DISCUSSION

We performed electrochemical etching with an EDLT configuration, as schematically drawn in the inset of Fig. 1(b). The upper panel of Fig. 1(a) shows the temporal change in the gate bias ($V_G$) and gate current ($I_G$) during etching. When applying a negative $V_G$, $I_G$ flowed from the SVO channel to the gate electrode, and anions with equal charges flowed from the gate electrode to the

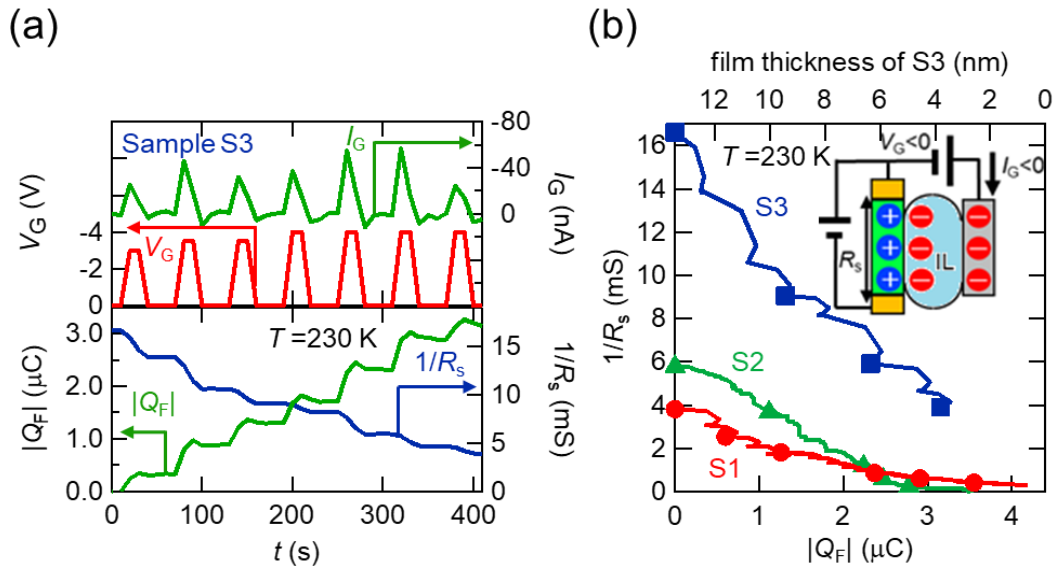

**FIG. 1.** (a) Time evolution of the gate bias ($V_G$), gate current ($I_G$), absolute value of the Faradaic charge ($|Q_F|$), and sheet conductance ($1/R_s$) during electrochemical etching at 230 K for sample S3. (b) $1/R_s$ vs. $|Q_F|$ curves for electrochemical etching processes at 230 K for three samples (S1-S3). The inset shows a schematic for an EDLT on an SVO film. $I_G$ flows to a gate electrode by applying $V_G$, and the anions in the ionic liquid (IL) move to the surface of the SVO film. The surface is etched by an electrochemical reaction.

SVO channel through the IL, resulting in an electrochemical etching reaction at the interface between the IL and SVO channel. The reacted amount of the SVO channel is proportional to the absolute value of the Faradaic charge ($Q_F$), where $Q_F$ is a temporal integration of $I_G$, $Q_F = \int I_G dt$. Therefore, $Q_F$ is proportional to a change in the thickness ($\Delta t$): $\Delta Q_F \propto \Delta t$. Thus, as shown in the lower panel of Fig. 1(a), the sheet conductance of the channel ($1/R_s$) decreased proportionally to $\Delta t$ and $\Delta|Q_F|$. Figure 1(b) shows $1/R_s$ as a function of $\Delta|Q_F|$ during etching processes for three samples (S1-S3). $1/R_s$ decreased to zero after 5-7 cycles of etching. The "final thickness" is the thickness of the film just before the final process. The film thicknesses after each cycle were estimated from the initial thicknesses, final thicknesses, and the relationship $\Delta Q_F \propto \Delta t$. We determined the final thickness based on the transport data obtained for each etched film. A detailed method is described in the Supplementary Information.

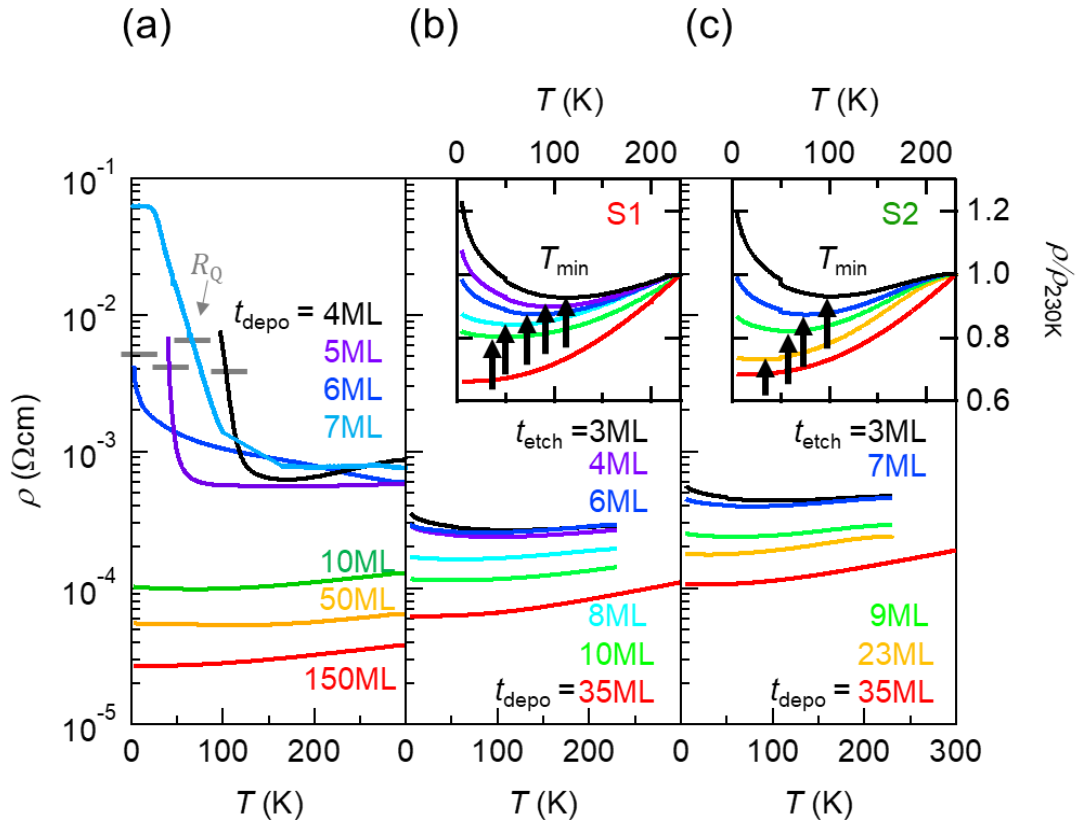

**FIG. 2.** a Temperature dependence of the resistivity ($\rho$) for the as-deposited SVO/STO films with thicknesses ($t_{depo}$) ranging from 4 to 150 monolayers (ML). The gray lines correspond to $\rho$, where $R_s$ is equal to the quantum resistance $R_Q = h/e^2 \sim 25$ k$\Omega$. b,c Temperature dependence of $\rho$ for the etched samples (b) S1 and (c) S2 with thicknesses ($t_{etch}$) ranging from 3 to 35 ML. The insets show $\rho$ normalized to that at 230 K ($\rho/\rho_{230\,K}$) as a function of temperature. Black arrows show the temperatures for the resistivity minima ($T_{min}$).

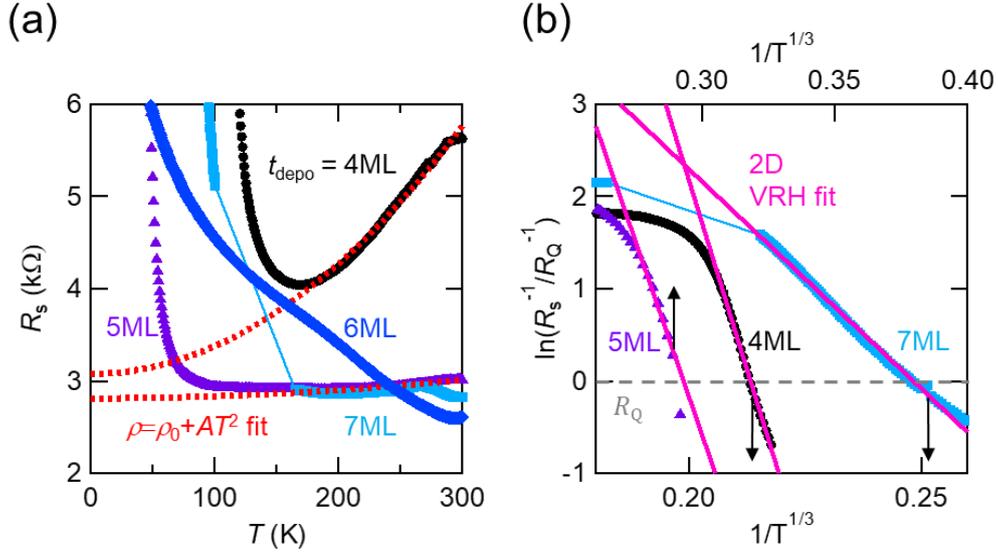

**FIG. 3.** (a) Temperature dependence of the sheet resistance ($R_s$) for as-deposited SVO films with $t_{depo}$ ranging from 4 to 7 ML. Fits to $\rho=\rho_0+AT^2$ are also shown as red dotted lines. (b) $\ln(R_s^{-1}/R_Q^{-1})$ as a function of $1/T^{1/3}$ for the as-deposited films with $t_{depo}$ ranging from 4 to 7 ML. Magenta straight lines correspond to fits of the data to the 2D variable range hopping (VRH) model. The gray dotted line marks the location of zero, where $R_s$ is equal to $R_Q$.

We examined the temperature dependence of $\rho$ for the etched SVO/LSAT films and as-deposited SVO/STO films with various thicknesses. All SVO films except for the as-deposited film with thickness ($t_{depo}$) = 7 ML showed an increase in $\rho$ near 300 K with decreasing film thickness. All of the etched films showed metallic behavior with resistivity minima at $T_{min}$, as shown in Figs. 2(b) and 2(c). The $T_{min}$ for the etched films shifted to higher temperatures with reduced film thickness. In contrast, Fig. 2(a) shows that the as-deposited films with $t_{depo}$ = 7 and 6 ML exhibited insulating behavior over the entire temperature range. Furthermore, the as-deposited films with $t_{depo}$ = 5 and 4 ML showed metallic behavior above 120 and 50 K, respectively, and showed a rapid increase in $\rho$ at lower temperatures. In the metallic regime, the temperature dependence of $\rho$ can be expressed by $\rho=\rho_0+AT^2$, corresponding to a Fermi liquid system, as shown in Fig. 3(a) by red dotted lines. The etched films also follow this relation in the metallic regime (a detailed discussion is given in Fig. 6).

Let us first focus on the temperature dependence of $R_s$ for the as-deposited SVO/STO films with $t_{depo}$ ranging from 4 to 7 ML. The as-deposited films with $t_{depo}$ = 4, 5, and 7 ML showed a crossing of the sheet resistance ($R_s$) with the quantum resistance $R_Q = h/e^2 \sim 25\text{k}\Omega$. In 2D, $R_s$ is related to $k_F l$ by

$$k_F l = \frac{R_Q}{R_S}, \quad (1)$$

where $l$ is the mean free path and $k_F$ is the Fermi wave vector. A system with $R_s$ above $h/e^2$ has $k_F l < 1$ and shows insulating behavior [15,29]. A previous study for semiconductors reported metallic and insulating behavior for samples with $R_s$ below and above $h/e^2$, respectively, over the entire temperature range [30]. This differs from our observation for the as-deposited SVO films. The crossing of $R_s$ with $h/e^2$ has also been reported in other strongly correlated systems [15,20,22]. This suggests that the insulating behavior in the ultrathin as-deposited SVO films is induced not only by a reduction in the film thickness itself but also by disorder in the films. This insulating behavior is described by the 2D Mott variable range hopping (2D VRH) model. In this model, $1/R_s$ is expressed as follows:

$$\frac{R_S^{-1}}{R_Q^{-1}} = C\exp\left[-\left(\frac{T_0}{T}\right)^{\frac{1}{3}}\right], \quad (2)$$

$$T_0 = \frac{13.8}{k_B N(E_F)\xi^2}, \quad (3)$$

where $k_B$ is the Boltzmann constant, $N(E_F)$ is the density of localized states at the Fermi level ($E_F$), and $\xi$ is the localization length. As shown in Fig. 3(b), $1/R_s$ is fitted by Eq. (2) near $R_Q$, where the values of $T_0$ are extracted from the slope to be $1.08\times10^5$, $5.86\times10^5$, and $4.42\times10^6$ K for the films with $t_{depo}$ = 7, 5, and 4 ML, respectively. According to VRH theory [20,31,32], the mean hopping distance ($l_{hop}$) is given by $l_{hop} = \left(\frac{T_0}{T}\right)^{\frac{1}{3}} \xi/3$, where $l_{hop}$ must be larger than $\xi$, namely, $l_{hop}/\xi > 1$. We estimated the value of $l_{hop}/\xi$ at characteristic temperatures $T_Q$, where $R_s(T_Q) = R_Q$, to be 3.94, 8.09, and 11.7 for the films with $t_{depo}$ = 7, 5, and 4 ML, respectively. All these values are larger than 1, which is consistent with VRH theory. By reducing $t_{depo}$ from 7 to 4 ML, $T_0$ is increased by more than one order of magnitude, and, thus, $N(E_F)\xi^2$ is decreased by the same order of magnitude according to Eq. (3). It is unlikely that $N(E_F)$ decrease by more than one order of magnitude by reducing $t_{depo}$ from 7 to 4 ML. Therefore, $\xi$ would be decreased by a factor of 6 or 7 by reducing $t_{depo}$; in other words, localization effects are enhanced by reducing the film thickness for the as-deposited ultrathin films.

Next, we examined the thickness dependence of the transport data for the etched and as-deposited films. Figure 4(a) shows plots of $\rho$ at 200 K ($\equiv \rho_{200K}$) against the film thickness on a double logarithmic scale. Red solid circles and green solid triangles represent the data for the etched films S1 and S2, respectively. For S3, only the available data at 230 K ($\rho_{230K}$) are plotted with blue solid squares. Open diamonds and squares indicate $\rho_{200K}$ for the as-deposited SVO/LSAT and SVO/STO films, respectively, where black diamonds and squares are collected from the data in references [17] and [24], respectively. Figure 4(b) displays $T_{min}$ versus the film thickness for the etched S1 and S2 films and as-deposited SVO/LSAT and SVO/STO films on a linear-linear scale, where the symbols are the same as in (a). For all the films except for S3, we found a trend in which $\rho_{200K}$ increased by one order of magnitude with decreasing film thickness from 50-10 ML to 6 ML. $T_{min}$ also showed a significant increase with decreasing film thickness, indicating the occurrence of insulating behavior at higher temperatures.

We interpret these results in terms of a metal to Mott-insulator transition being associated with the reduction of the film thickness (reduced dimensionality). According to the Hubbard model, a system shows a transition from metallic to Mott insulator states with decreasing bandwidth ($W$), and it is known that $W$ in a 2D system is smaller than that in a 3D system. In this framework, therefore, the observed increase in $\rho_{200K}$ and $T_{min}$ with decreasing film thickness is attributed to the change in the electronic states from metallic to Mott insulator states due to the reduced $W$.

For both the as-deposited and etched films shown in Fig. 4(a), we observed a common trend for $\rho_{200K}$, which increased with decreasing film thickness. On the other hand, the value of $\rho_{200K}$ for a given thickness differs largely (which spans a factor of 3-5), depending on the series of SVO films. In particular, for the series of as-deposited films, it is even difficult to draw individual smooth lines of a similar shape on which data points fall. For the S1 and S2 etched films, in contrast, all the data points in each series fall on nearly smooth lines of similar shape.

The difference in $\rho_{200K}$ for a given thickness observed among the series of etched and as-deposited films can result from the difference in disorder in the films. The large scattering of data points for the as-deposited films is reasonable considering that each sample with different $t_{depo}$ was prepared independently; hence, the disorder is dependent on the preparation conditions. In contrast, for the etched films, the disorder in the series of films with different $t_{etch}$ is determined mainly by that in the initial film before etching, and we can expect that the disorder is not significantly changed during etching. This is why a systematic change in $\rho_{200K}$ was observed for the etched films with a reduced film thickness.

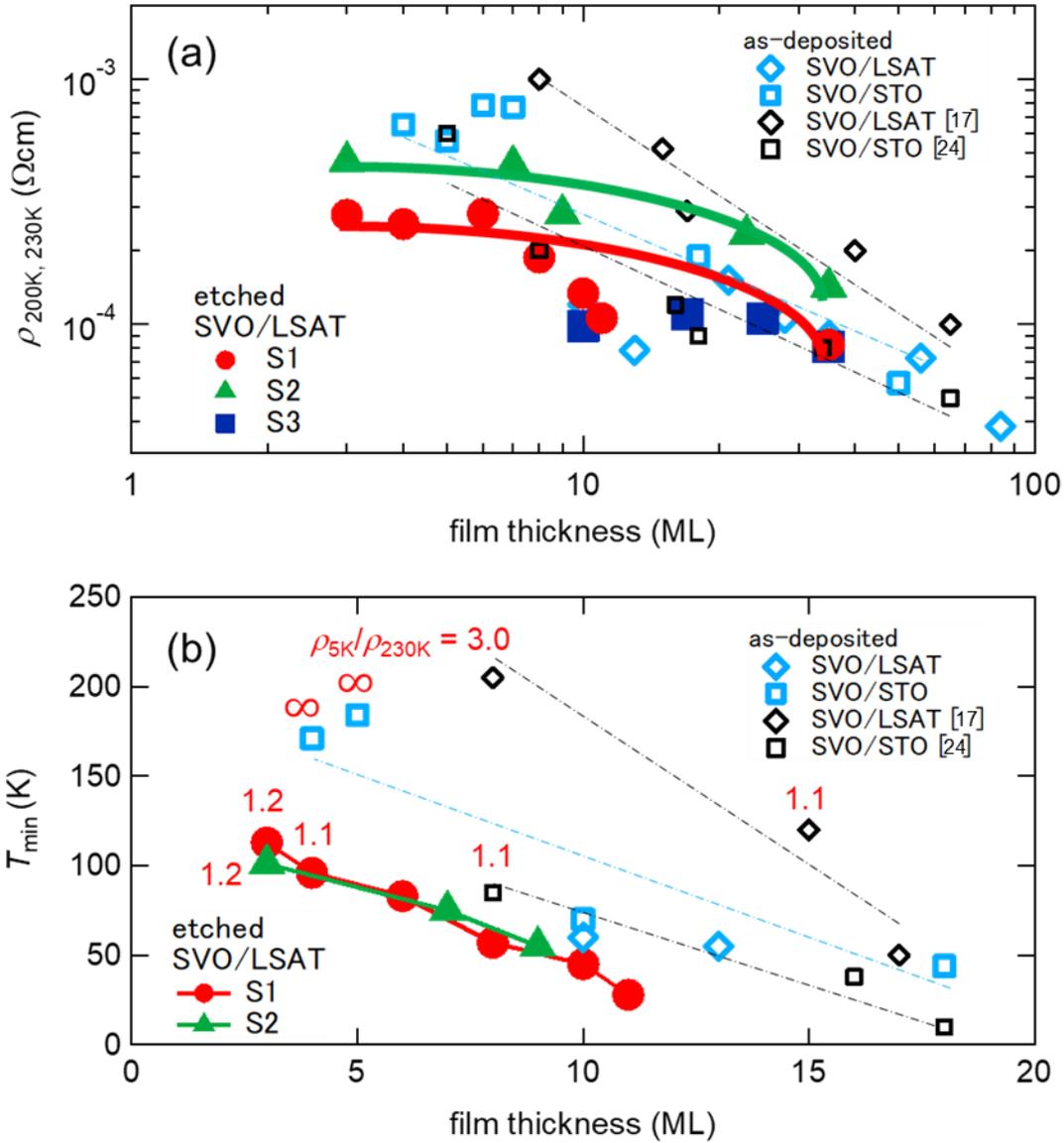

**FIG. 4.** (a) $\rho$ at 200 K ($\rho_{200\,K}$) plotted against the film thickness on a log-log scale for S1 (red solid circles) and S2 (green solid triangles). For S3, $\rho$ at 230 K ($\rho_{230K}$) is plotted with blue solid squares. Open symbols represent $\rho_{200K}$ for as-deposited films (SVO/LSAT and SVO/STO), where the data denoted by black diamonds and squares are taken from references [17] and [24]. All lines are guides to the eye. (b) $T_{min}$ as a function of the film thickness on a linear scale for S1 and S2 and as-deposited films (SVO/LSAT and SVO/STO). The symbols are the same as in (a). The values of $\rho_{5K}/\rho_{230K}$ are also shown for the films with $\rho_{5K}/\rho_{230K} > 1$. We denote $\rho_{5K}/\rho_{230K}$ as $\infty$ when $\rho$ exhibits an abrupt increase by more than one order of magnitude above 50 K upon cooling. All lines are guides to the eye.

This view is further supported by the following fact: The data points for $T_{min}$ versus thickness for S1 and S2 fall on a single line, in contrast to the behavior observed for the as-deposited films. Furthermore, the $T_{min}$ versus thickness line for the etched S1 and S2 films indicated with full lines is located below the $T_{min}$ versus thickness line for as-deposited films indicated by dashed lines. These features strongly suggest that the etched films have small disorder, which is introduced before etching, and its strength stays nearly constant even after the films are etched down to 3 ML. The small disorder in the S1 and S2 etched films explains the weakly insulating behavior observed at low temperature down to the ultrathin regime, as shown in Figs. 2(b) and 2(c).

To heuristically quantify the strength of the localization effects at low temperature in each film, we focus on the ratio of $\rho$ at the lowest temperature to that at high temperature, $\rho_{5K}/\rho_{230K}$. Then, we defined the films with $\rho_{5K}/\rho_{230K} > 1$ as "insulating" and the other films as "weakly insulating". In Fig. 4(b), we show the values of $\rho_{5K}/\rho_{230K}$ for the "insulating" films with $\rho_{5K}/\rho_{230K} > 1$. Upon cooling, $\rho$ for the as-deposited films with $t_{depo} < 7$ ML showed an abrupt increase by more than one order of magnitude above 50 K. Hence, we denote $\rho_{5K}/\rho_{230K}$ as $\infty$. It is notable that even for the 15 ML thick film, the as-deposited film is "insulating" [17].

Similar "insulating" behaviors with thicknesses of approximately 10 ML have been reported in transition metal oxide thin films, such as $CaVO_3$ [16], $NdNiO_3$ [19], and $LaNiO_3$ [14]. In addition, an abrupt increase in $\rho$, corresponding to a $\rho_{5K}/\rho_{230K}$ value of $\infty$, at low temperature has also been reported for ultrathin films of $La_{2/3}Sr_{1/3}MnO_3$ [18] and $SrRuO_3$ [22]. These insulating behaviors have been considered to be due to various disorder effects in thin films, such as strain effects due to the substrate, oxygen vacancies, and surface effects [18,19]. In the case of the SVO films, the etched films are coated by IL; this is in contrast to the as-deposited films, which are exposed to air. Therefore, the "insulating" behavior of the as-deposited SVO films is most likely induced by surface disorder effects. In contrast, for the S1 and S2 etched films, the films with thicknesses above 6 ML were "weakly insulating", and only the 3- and 4-ML-thick films were "insulating". Moreover, the value of $\rho_{5K}/\rho_{230K}$ for the thinnest (3 ML) films is at most 1.2. These results indicate that the etched films are robust against disorder. A previous study using *in situ* photoemission (PES) showed that Mott insulating states are induced in ultrathin SVO films with a thickness lower than 3 ML [23]. Thus, we conclude that the SVO ultrathin films fabricated by electrochemical etching have very low disorder effects comparable to an *in situ* SVO film.

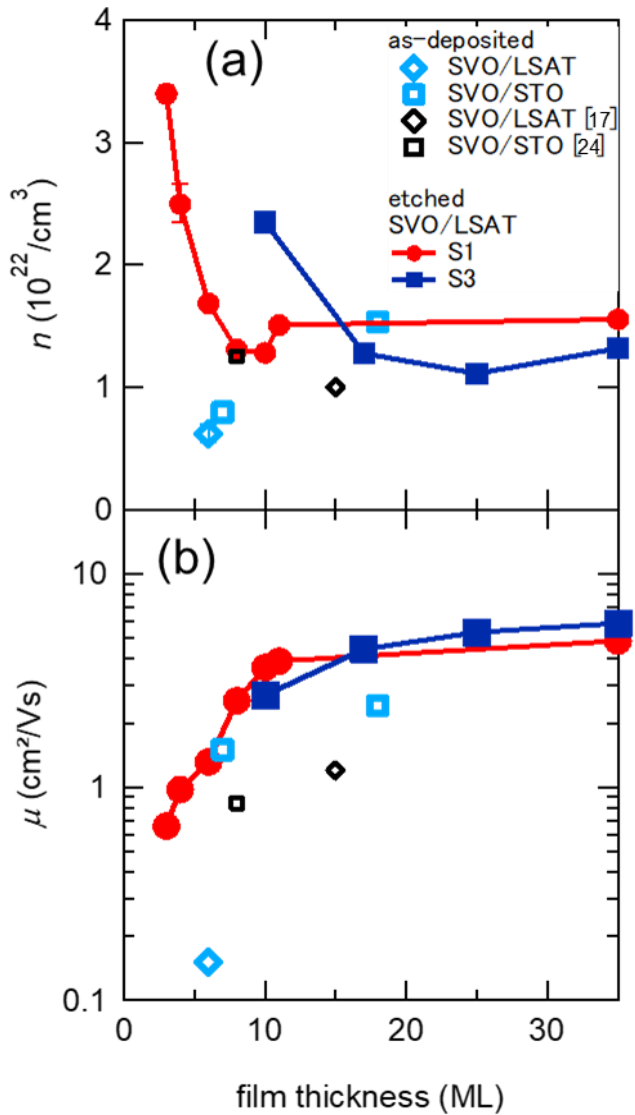

**FIG. 5.** (a) Carrier density (*n*) and (b) mobility ($\mu$) as a function of the film thickness at 200 K for S1 (red solid circles) and at 230 K for S3 (blue solid squares). Data for the as-deposited films (SVO/LSAT at 300 K and SVO/STO at 100 K) and those (SVO/LSAT and SVO/STO at 200 K) obtained from the literature are also plotted [17,24].

The carrier density ($n$) and electron mobility ($\mu$) were estimated from Hall measurements for the etched films and as-deposited films. As shown in Fig. 5(a), for all the films, $n$ was almost independent of the film thickness above 10 ML, suggesting that thicker films above 10 ML are bulk-like. The S1 and S3 etched films showed an increase in $n$ with decreasing film thickness below approximately 10 ML, as shown with red solid circles and blue solid squares, respectively, contrasting with a decrease in $n$ for the as-deposited films shown with open symbols. The decrease in $n$ observed for the as-deposited films most likely originates from increased disorder for the thinner films, such as surface effects and defects at the interface between the film and the substrate. The increase in $n$ observed for the etched films might originate from unexpected intercalation of ions into the film or adsorption of ions on the film during the electrochemical etching process.

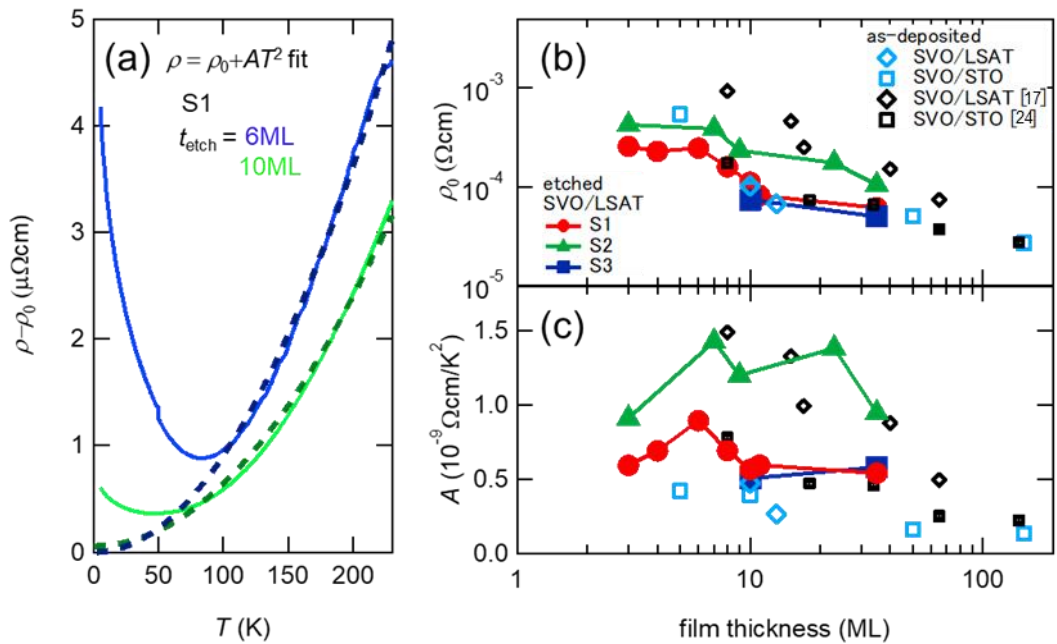

**FIG. 6.** (a) $\rho-\rho_0$ as a function of $T$ for S1 with $t_{etch}$ = 6 ML (blue) and 10 ML (green). Dashed lines represent the $\rho=\rho_0+AT^2$ fits. (b) The fitted parameters $\rho_0$ and (c) $A$ for S1 (red solid circles), S2 (green solid triangles), and S3 (blue solid squares) plotted against the film thickness. Open symbols represent the data for the as-deposited films (SVO/LSAT and SVO/STO), where black diamonds and squares denote data obtained from references [17] and [24], respectively.

As shown in Fig. 5(b), the data points for $\mu$ for the S1 and S3 etched films fall on nearly the same smooth curve that shows a decrease with reducing film thickness below approximately 10 ML. The as-deposited films also showed a decreasing trend for $\mu$ with decreasing film thickness; however, $\mu$ for the as-deposited films with the same thicknesses differed from each other. This indicates that each as-deposited film had different disorder effects. The monotonic decrease in $\mu$ with decreasing film thickness can result from two origins: first, an increase in disorder around the interface between the film and the substrate, and, secondly, from an enhancement of the electron effective mass, which is expected for a system approaching the MIT with decreasing dimensionality.

According to the Fermi liquid theory, the temperature dependence of $\rho$ in a Fermi liquid system is expressed by $\rho=\rho_0+AT^2$, where the residual resistivity $\rho_0$ is attributed to static disorder [33,34] and the parameter $A$ is related to an electron effective mass enhanced by an electron-electron interaction in a strongly correlated metal [10]. In our experiment, $\rho$-$T$ curves with various thicknesses follow that relationship at high temperatures above $T_{min}$. Typical results for the fitting are shown for S1 films with $t_{etch}$ = 6 and 10 ML by dashed lines in Fig. 6(a), together with the $\rho$–$\rho_0$ versus $T$ data indicated by blue and green lines, respectively.

Figures 6(b) and 6(c) display the parameters $\rho_0$ and $A$, respectively, extracted from the fitting as a function of the film thickness for S1 (red solid circles), S2 (green solid triangles), and S3 (blue solid squares). Open symbols in Figs. 6(b) and 6(c) denote the parameters $\rho_0$ and $A$, respectively, for the as-deposited films (SVO/LSAT and SVO/STO), where black diamonds and squares denote data obtained from references [17] and [24], respectively. Figure 6(b) shows that for all the films, the dependence of $\rho_0$ on the film thickness is nearly the same as that of $\rho_{200K}$ (or $\rho_{230K}$) on the film thickness, indicating that $\rho_0$ and $\rho_{200K}$ (or $\rho_{230K}$) reflect essentially the same properties related to static disorder. Figure 6(c) shows that for both etched and as-deposited films, $A$ shows a roughly increasing trend with decreasing film thickness from approximately 30-100 ML down to 6-7 ML. For S1 and S2, in particular, as $t_{etch}$ is reduced below approximately 10 ML, after showing a small peak at approximately 6-7 ML, $A$ exhibits a decrease in the thinnest region of 3-6 ML.

According to Ref. [35,36], $A$ is expressed by the following equation:

$$A = \left(\frac{4\pi^2 k_B^2}{e^2\hbar^2}\right)\left(\frac{m_b}{n}\right)\Phi, \quad (4)$$

where $k_B$ is the Boltzmann constant, $e$ is the elementary charge, $\hbar$ is the Plank constant divided by $2\pi$, $m_b$ is the band mass, and $\Phi$ is the coefficient of the scattering rate. An enhancement of electron-electron interactions leads to an increase in $m_b\Phi$, which, in turn, results in an increase in A, in accordance with Eq. (4). Equation (4) also indicates that an increase in n leads to a reduction in A. Indeed, a decrease in A associated with an increase in n has been reported for SrTi$_{1-x}$V$_x$O$_3$ near the MIT [37,38].

The increase in A observed below 10 ML down to 6-7 ML for the etched films is attributed to an enhanced $m_b\Phi$ due to enhanced electron-correlation effects induced by the reduced film thickness. Here, we consider the experimental findings that as the film thickness decreases from 10 to 6-7 ML, the available data for n for S1 exhibit a slight decrease, as shown in Fig 5(a), which gives rise to an increase in A. In contrast, the decrease in A observed below 6-7 ML cannot be explained by the enhanced electron-correlation effects mentioned above. We consider that this is due to the competing effects between a pronounced increase in n shown in Fig. 5(a) and a moderate increase in $m_b\Phi$ due to the enhanced electron correlation. The former effect is more dominant than the latter.

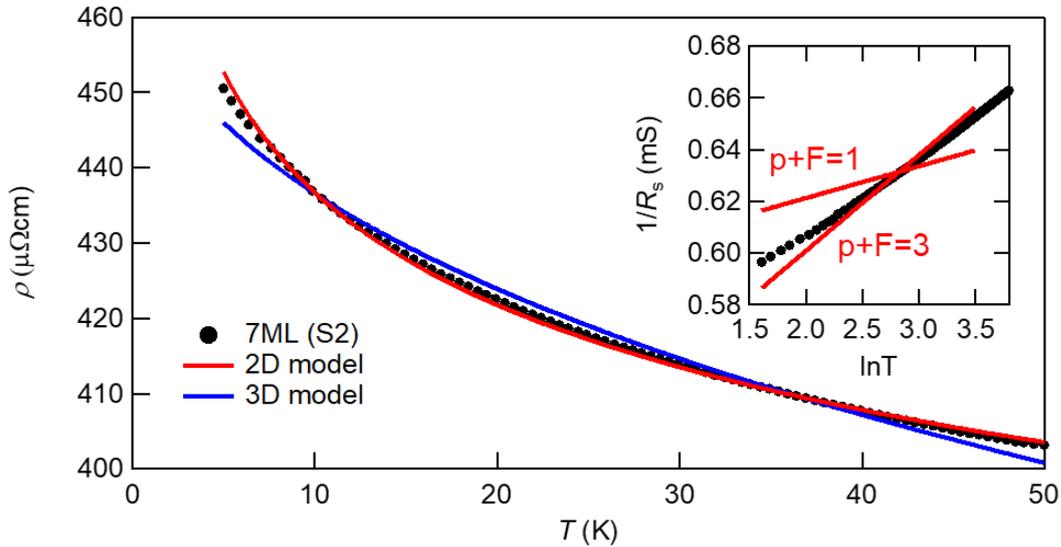

**FIG. 7.** $\rho$-T data for the 7 ML-thick S2 film. Blue and red lines represent fits of the data to Eq. (6) (3D QCC model) and Eq. (7) (2D QCC model), respectively. The inset shows 1/$R_s$ versus the logarithm of temperature for the same data. Two straight red lines with small and large slopes correspond to the fits to Eq. (7) with p+F=1 and 3, respectively.

All of the etched SVO films exhibit metallic behavior with a resistivity upturn at low temperature. We examine the origin of the resistivity upturn in the etched films. Near the metal-insulator transition, electrons are weakly localized, and the conductivity is affected by electron-electron interactions and interference. In this regime, the resistivity upturn has been previously interpreted using the concept of quantum corrections $\Delta\sigma(T)$ to the conductivity $\sigma(T)$ (QCC) [39,40]. The QCC is explained by two contributions: weak localization (WL) and a renormalized electron-electron interaction (REEI). We define $\Delta\sigma(T)$ as a correction to the Fermi liquid model. Then, $\rho$ is expressed as follows:

$$\rho = \frac{1}{\sigma_0 + \Delta\sigma(T)} + AT^2, \quad (5)$$

where $\sigma_0$ is the residual conductivity and $A$ is the parameter in the Fermi liquid model mentioned above. The WL is reflected in a resistivity correction due to electron interference scattering by defects or impurities [39,40], and the REEI originates from the density of state correction at the Fermi level.

In the 3D and 2D QCC models, $\Delta\sigma$ is expressed as follows:

$$\Delta\sigma_{3D} = a_1 T^{\frac{q}{2}} + a_2 T^{\frac{1}{2}}, \quad (6)$$

$$\Delta\sigma_{2D} = a\ln T = \frac{e^2}{\pi h}(p + F)\ln T, \quad (7)$$

where $a_1, q$ and $p$ arise from WL, and $a_2$ and $F$ correspond to the contribution from REEI. When electron-electron scattering governs the conductivity, the parameters $q$ and $p$ are equal to 2 and 1, respectively, while $q$ and $p$ are both equal to 3 for electron-phonon scattering. In addition, the correction due to the WL is always positive, so the parameters $a_1$ and $p$ are positive.

We fitted the $\rho$-$T$ data for various thicknesses using Eqs. (6) and (7) with a fixed $q = 3$. Figure 7 shows typical results for the fitting of $\rho(T)$ for the 7 ML-thick S2 film. The 2D fit (red line) follows the experimental data better than the 3D fit (blue line). This indicates that the 2D QCC model is better applied to describe $\rho(T)$ for the etched films. The parameter $p+F$ in Eq. (7) can be extracted from the slope of $1/R_s$ versus the logarithm of temperature, as shown in the inset of Fig. 7. The $p+F$ values for the S1 films with $t_{etch}$ = 10, 8, 6, 4, and 3 ML are 2.8, 2.7, 3.0, 3.0, and 2.8, respectively. For the S2 films with $t_{etch}$ = 9, 7, and 3 ML, the $p+F$ values are 2.8, 2.4, and 1.5, respectively. Thus, all these $p+F$ values turn out to range between 1 and 3. Although we are unable to specify each value of $p$ and $F$ separately, considering the theoretical prediction that $p$ is either 1 or 3, the present result strongly suggests that $F$ is at most of the order of 0.1, i.e., the contribution of WL is more important than that of REEI. Indeed, it has been reported that $F$ ranges from 0.34 to 0.57 in Ref. [41]. As mentioned above, for all the etched films except for the 3 ML-thick S2 film, the values of $p+F$ remain at approximately 3. This finding indicates that electron-phonon scattering is the dominant scattering mechanism rather than electron-electron scattering.

We measured $\rho$ for S2 films with different film thicknesses as a function of the perpendicular magnetic field ($B$) at 5 K. Figure 8(a) shows the normalized magnetoresistance (MR), which is defined as $100 \times (\rho(B) - \rho(0))/\rho(0)$. All S2 films with thicknesses ranging from 3 to 23 ML showed a positive MR. For some semiconductors, a positive MR with a parabolic shape, $\mathrm{MR} \approx (\mu B)^2$, was reported, whose origin is attributed to a Lorenz force. However, in our case, $(\mu B)^2$ is estimated to be as small as $10^{-5}\%$ for 3 T, which is three orders of magnitude smaller than the observed value.

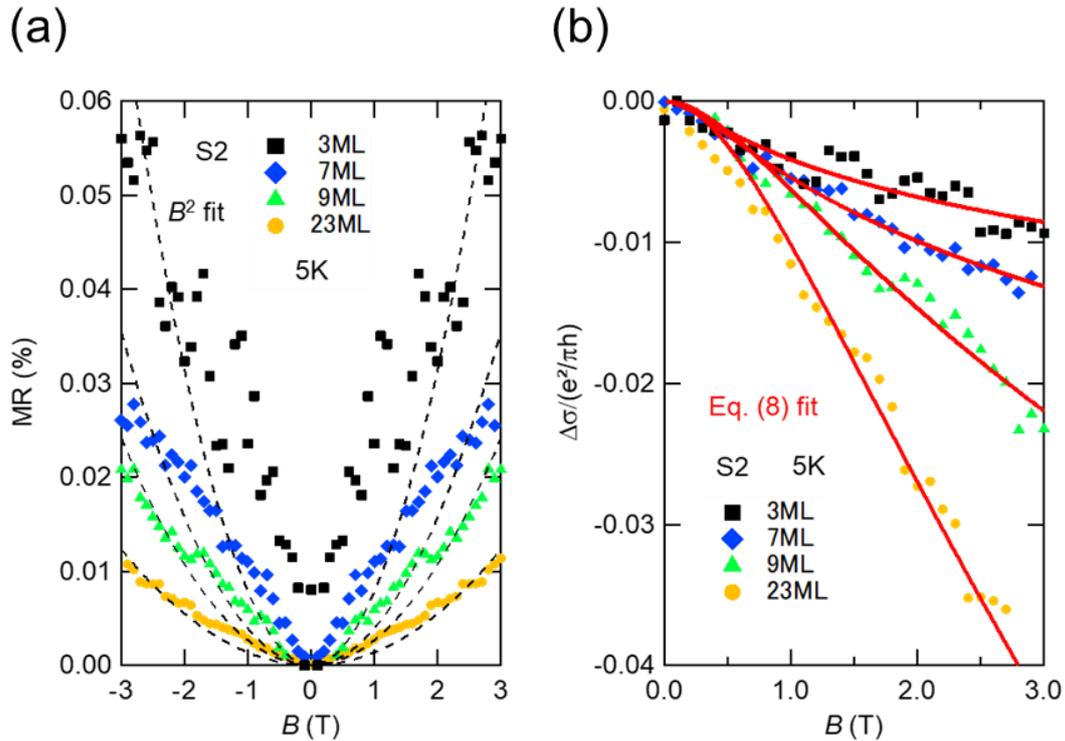

**FIG. 8.** (a) Perpendicular magnetic field ($B$) dependence of the normalized magnetoresistance (MR), $100 \times (\rho(B) - \rho(0))/\rho(0)$, for S2 with thicknesses ranging from 3 to 23 ML at 5 K, where $\rho(0)$ is the resistivity minimum in the MR curve. Dashed lines are $B^2$ fits. (b) Magnetoconductance normalized by the quantum conductance($e^2/\pi h$) as a function of $B$ for samples with thicknesses ranging from 3 to 23 ML. Red lines are fits to the data based on Eq. (8).

We show, instead, that our positive MR is well explained in terms of the weak anti-localization (WAL) effect for a weakly disordered metal [24,42-44]. The magnetoconductance is well described by the Hikami-Larkin-Nagaoka (HLN) equation in 2D, where the spin-orbit length, $l_{so}$, is negligibly smaller than the inelastic scattering length, $l_{in}$ ($l_{so} \ll l_{in}$) [42].

$$(\sigma(B) - \sigma(0))/(\tfrac{e^2}{\pi h}) = \alpha[\ln\left(\tfrac{B_{in}}{B}\right) - \psi(\tfrac{1}{2} + \tfrac{B_{in}}{B})], \quad (8)$$

$$B_{in} = \frac{h}{8\pi l_{in}^2}, \quad (9)$$

where $\psi$ is the digamma function, $B_{in}$ is a characteristic magnetic field for inelastic scattering, and $l_{in}$ is the inelastic mean free path. Figure 8(b) shows the magnetoconductance converted from the MR data in Fig. 8(a) and fits to the data based on the HLN equation (Eq. (8)). The fitting parameter $\alpha$ is an indication of the dominant effect: $\alpha = -1$ for WL and $0 < \alpha < 1/2$ for WAL. For films with thickness of 23 ML (~8.8 nm), 9 ML, 7 ML, and 3 ML, $\alpha$ is determined to be 0.098, 0.034, 0.011, and 0.0053, and $l_{in}$ is determined to be 17, 21, 32, and 41 nm, respectively. $l_{in}$ is always larger than the film thickness, and, thus, it is confirmed that the etched films are in the 2D regime.

Interestingly, competition between WAL and WL effects has been reported in the weakly localized regime above 5 K for as-deposited SVO films with thicknesses of 4 and 5 ML [21]. This indicates that $l_{so}$ is of the same order of magnitude as $l_{in}$ for the films in Ref. [21]. It is considered that $l_{so}$ for our etched SVO films is almost the same as that for the as-deposited SVO films in Ref. [21] because $l_{so}$ is determined by the spread of the *d*-orbital of vanadium, namely, it only depends on the compound. Therefore, $l_{in}$ for our films is larger than in Ref. [21], indicating that the disorder effects for our etched films are smaller than those for the as-deposited films in Ref. [21].

## CONCLUSION

We fabricated conductive ultrathin SVO films by electrochemical etching, which exhibited metallic behavior down to 3 ML. The films showed a systematic change in transport properties with decreasing film thickness, indicating that the disorder remained nearly unchanged during etching and that only the thickness was reduced. In contrast, the as-deposited ultrathin SVO films behaved like a strongly localized metal, which was attributed to disorder introduced during deposition. The resistivity $\rho$ (*T*) was observed to show insulating behavior for the as-deposited ultrathin films below 10 ML, which was reproduced by a 2D VRH model, in which the localization length decreased with decreasing film thickness.

For the etched films, the electron mobility measured at 200 K showed a decrease with decreasing film thickness below 10 ML. This was interpreted as originating from an increased scattering rate and an enhanced electron effective mass, as expected for a system approaching a MIT with reduced dimensionality. The etched films and most of the as-deposited films showed Fermi liquid behavior in the high-temperature regime. In the low-temperature regime, a slight upturn in $\rho$ (*T*) and a positive MR were observed for the etched films down to 3 ML, which was explained in terms of weak anti-localization theory for a weakly disordered metal.

## ACKNOWLEDGMENTS

This work was supported in part by JSPS KAKENHI (Grant Numbers 21H01038 and 19H02798) and CREST-JST (Grant Numbers JPMJCR15Q2).

## REFERENCES

[1] N. F. Mott, Proc. Phys. Soc. London, Sect. A, **62**, 416 (1949).

[2] N. F. Mott, Metal‐Insulator Transitions, 2nd ed. Taylor and Francis, London, (1990).

[3] P. W. Anderson, Phys. Rev. **109**, 1492 (1958).

[4] M. Imada, A. Fujimori, and Y. Tokura, Rev. Mod. Phys. **70**, 1039 (1998).

[5] F. Inaba, T. Arima, T. Ishikawa, T. Katsufuji, and Y. Tokura, Phys. Rev. B **52**, R2221 (1995).

[6] S. Miyasaka, T. Okuda, and Y. Tokura, Phys. Rev. Lett. **85**, 5388 (2000).

[7] T. M. Dao, P. S. Mondal, Y. Takamura, E. Arenholz, and J. Lee, Appl. Phys. Lett. **99**, 112111 (2011).

[8]  I. H. Inoue, I. Hase, Y. Aiura, A. Fujimori, Y. Haruyama, T. Maruyama, and Y. Nishihara, Phys. Rev. Lett. **74**, 2539 (1995).

[9]  M. J. Rozenberg, I. H. Inoue, H. Makino, F. Iga, and Y. Nishihara, Phys. Rev. Lett. **76**, 4781 (1996).

[10] I. H. Inoue, O. Goto, H. Makino, N. E. Hussey, and M. Ishikawa, Phys. Rev. B **58**, 4372 (1998).

[11] M. Takayanagi, T. Tsuchiya, W. Namiki, S. Ueda, M. Minohara, K. Horiba, H. Kumigashira, K. Terabe, and T. Higuchi, Appl. Phys. Lett. **112**, 133106 (2018).

[12] H. Y. Wang, Y. Iwasa, M. Kawasaki, B. Keimer, N. Nagaosa, and Y. Tokura, Nat. Mater. **11**, 103 (2012).

[13] R. Ramesh and D. G. Schlom, Nat. Rev. Mater. **4**, 257 (2019).

[14] J. Son, P. Moetakef, J. M. LeBeau, D. Ouellette, L. Balents, S. J. Allen, and S. Stemmer, Appl. Phys. Lett. **96**, 062114 (2010).

[15] R. Scherwitzl, S. Gariglio, M. Gabay, P. Zubko, M. Gibert, and J. M. Triscone, Phys. Rev. Lett. **106**, 246403 (2011).

[16] M. Gu, J. Laverock, B. Chen, K. E. Smith, S. A. Wolf, and J. Lu, J. Appl. Phys. **113**, 133704 (2013).

[17] M. Gu, S. A. Wolf, and J. Lu, Adv. Mater. **1**, 1300126 (2014).

[18] Z. Liao, F. Li, P. Gao, L. Li, J. Guo, X. Pan, R. Jin, E. W. Plummer, and J. Zhang, Phys. Rev. B **92**, 125123 (2015).

[19] L. Wang, S. Ju, L. You, Y. Qi, Y. W. Guo, P. Ren, Y. Zhou, and J. Wang, Sci. Rep. **5**. 18707 (2016).

[20] L. Wang, L. Chang, X. Yin, A. Rusydi, L. You, Y. Zhou, L. Fang, and J. Wang, J. Phys. Condens. Matter **29**, 025002 (2017).

[21] G. Wang, Z. Wang, M. Meng, M. Saghayezhian, L. Chen, C. Chen, H. Guo, Y. Zhu, E. W. Plummer, and J. Zhang, Phys. Rev. B **100**, 155114 (2019).

[22] X. Shen, X. Qiu, D. Su, S. Zhou, A. Li, and D. Wu, J. Appl. Phys. **117**, 015307 (2015).

[23] K. Yoshimatsu, T. Okabe, H. Kumigashira, S. Okamoto, S. Aizaki, A. Fujimori, and M. Oshima, Phys. Rev. Lett. **104**, 147601 (2010).

[24] A. Fouchet, M. Allain, B. Berini, E. Popova, P. E. Janolin, N. Guiblin, E. Chikoidze, J. Scola, D. Hrabovsky, Y. Dumont, and N. Keller, Mater. Sci.

Engineering. B **212,** 7 (2016).

[25] J. Shiogai, Y. Ito, T. Mitsuhashi, T. Nojima, and A. Tsukazaki, Nat. Phys. **12**, 42 (2016).

[26] M. Yoshida, J. Ye, T. Nishizaki, N. Kobayashi, and Y. Iwasa, Appl. Phys. Lett. **108**, 202602 (2016).

[27] S. Kouno, Y. Sato, Y. Katayama, A. Ichinose, D. Asami, F. Nabeshima, Y. Imai, A. Maeda, and K. Ueno, Sci. Rep. **8**, 14731 (2018).

[28] N. Shikama, Y. Sakishita, F. Nabeshima, Y. Katayama, K. Ueno, and A. Maeda, Appl. Phys. Express. **13**, 083006 (2020).

[29] D. C. Licciardello and D. J. Thouless, Phys. Rev. Lett. **35**, 1475 (1975).

[30] M. Nakano, A. Tsukazaki, A. Ohtomo, K. Ueno, S. Akasaka, H. Yuji, K. Nakahara, T. Fukumura, and M. Kawasaki, Adv. Mater. **22**, 876 (2010).

[31] N. F. Mott and E. A. Davis, Electronic Processes in Non-Crystalline Materials (Oxford Classic Texts in the Physical Sciences) (1970).

[32] Z. H. Khan, M. Husain, T. P. Perng, N. Salah, and S. Habib, J. Phys. Condens. Matter. **20**, 475207 (2008).

[33] I. H. Inoue, H. Makino, I. Hase, M. Ishikawa, N. E. Hussey, and M. J. Rozenberg, Phys. B. Condens. Matter. **237**, 61 (1997).

[34] Y. C. Lan, X. L. Chen, and M. He, J. Alloys Compd. **354**, 95 (2003).

[35] N. E. Hussey, J. Phys. Soc. Jpn. **74**, 1107 (2005).

[36] D. Oka, Y. Hirose, S. Nakao, T. Fukumura, and T. Hasegawa, Phys. Rev. B **92**, 205102 (2015).

[37] K. Hong, S. H. Kim, Y. J. Heo, and Y. U. Kwon, Solid. State. Commun. **123**, 305 (2002).

[38] M. Gu, S. A. Wolf, and J. Lu, Appl. Phys. Lett. **103**, 223110 (2013).

[39] G. Bergmann, Phys. Rep. **107**, 1 (1984).

[40] P. A. Lee and T. V. Ramakrishnan, Rev. Mod. Phys. **57**, 287 (1985).

[41] S. Mukhopadhyay and I. Das, J. Phys. Condens. Matter. **21**, 186004 (2009).

[42] S. Hikami, A.I. Larkin, and Y. Nagaoka, Prog. Theor. Phys. **63**, 707 (1980).

[43] A. D. Caviglia, M. Gabay, S. Gariglio, N. Reyren, C. Cancellieri, and J. M. Triscone, Phys. Rev. Lett. **104**, 126803 (2010).

[44] M. Liu, J. Zhang, C. Z. Chang, Z. Zhang, X. Feng, K. Li, K. He, L. l. Wang,

X. Chen, X. Dai, Z. Fang, Q. K. Xue, X. Ma, and Y. Wang, Phys. Rev. Lett. **108**, 036805 (2012).

# Supplementary Information
# Transport properties around the metal-insulator transition for SrVO$_3$ ultrathin films fabricated by electrochemical etching

Hikaru Okuma, Yumiko Katayama, Keisuke Otomo, and Kazunori Ueno
Department of Basic Science, University of Tokyo
Meguro, Tokyo 153-8902, Japan

## Methods used for determining the film thickness of the etched films

As mentioned in the Results and Discussion section, the thickness of the film after each cycle of etching, $t_{etch}$, was determined from the initial thickness of the as-deposited film before etching and the final thickness just before the final etching process, $t_{final}$, using the relation $\Delta Q_F \propto \Delta t$. While the initial thickness of 35 ML was directly obtained from the experiment, $t_{final}$ was estimated from the methods detailed below. We determined the final film thickness $t_{final}$ from the data for S1 and S2 without using the data for S3, because the final etching process was carried out for S1 and S2 but not for S3.

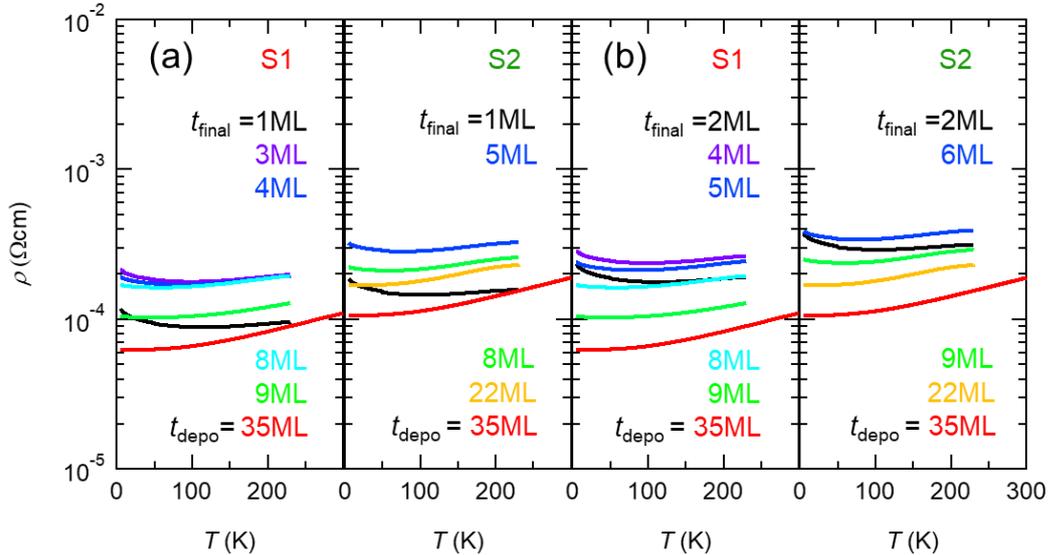

**FIG. SI-1.** $\rho$-$T$ curves for the etched films S1 and S2 with tentative thicknesses shown in the figure; these thicknesses were determined assuming that the final thicknesses $t_{final}$ are (a) 1 ML and (b) 2 ML. The left and right panels in each figure represent the data for the S1 and S2 films, respectively. The data for the S1 and S2 films with a thickness of $t_{depo}$ = 35 ML before etching are also shown with red curves.

As shown in Fig. 1(b) in the main text, the sheet conductance $1/R_s$ almost linearly decreased with increasing $|Q_F|$, and reached close to zero. Indeed, the ratio between the final and the 35 ML thickness, $\frac{1/R_s(t_{\text{final}})}{1/R_s(35\text{ML})}$ is calculated to be 2.8 % and 4.3 % for S1 and S2, respectively. Thus, in the first approximation, we expect that a simple linear relation $\frac{t_{\text{final}}}{35\text{ML}} = \frac{1/R_s(t_{\text{final}})}{1/R_s(35\text{ML})}$ holds in the entire thickness region ranging from 35 ML to $t_{\text{final}}$. This gives $t_{\text{final}}$ = 1.0 and 1.5 ML for S1 and S2, respectively. Hence, it is expected that the actual values of $t_{\text{final}}$ are a few monolayers and the difference in $t_{\text{final}}$ between S1 and S2 is smaller than 1 ML. Therefore, in the first approximation, we consider that the values of $t_{\text{final}}$ for S1 and S2 are identical.

Next, we assumed the final film thicknesses $t_{\text{final}}$ to be 1 ML or 2 ML, and $\rho$-$T$ curves were plotted with tentative thicknesses, as shown in Figs. SI-1(a) ($t_{\text{final}}$ =1ML) and 1(b) ($t_{\text{final}}$ =1ML). The $\rho$-$T$ curves for the S1 and S2 films with the thickness of $t_{\text{depo}}$ = 35 ML before etching are also shown with red curves. As shown in Fig. SI-1(a), $\rho$ for the S1 and S2 films with $t_{\text{final}}$ = 1 ML are significantly lower than those for the S1 and S2 films with $t_{\text{etch}}$ = 8 ML, respectively, over the entire temperature region. Similarly, Fig. SI-1(b) shows that the $\rho$ for the S1 and S2 films with $t_{\text{final}}$ = 2 ML are lower than those for the S1 and S2 films with $t_{\text{etch}}$ = 5 ML and 6 ML, respectively, in the entire $T$ region. This behavior is physically unusual, indicating that the assumed values for $t_{\text{final}}$ = 1 and 2 ML are not appropriate.

The $\rho/\rho_{230K}$-$T$ curves for the S1 and S2 films with the final thickness $t_{\text{final}}$ nearly coincide with each other, as shown in the insets of Figs. 2(b) and 2(c) in the main text, respectively. Hence, $T_{\text{min}}$ and $\rho_{5K}/\rho_{230K}$ ($\rho$ at 5 K normalized to that at 230 K) with the final thickness for S1 are also in accordance with those for S2, respectively. In addition, The $\rho/\rho_{230K}$-$T$ curves for the S1 and S2 films with an initial thickness of 35 ML also coincide with each other. Then, we assume that a $\rho/\rho_{230K}$-$T$ curve for S1 would follow a similar curve to that for S2 when their thickness is same. We employ $T_{\text{min}}$ and $\rho_{5K}/\rho_{230K}$ as useful indicators to detect the similarity of the $\rho/\rho_{230K}$-$T$ curve. In the following analysis, we will estimate an actual value of $t_{\text{final}}$ based on that thickness dependences of $T_{\text{min}}$ and $\rho_{5K}/\rho_{230K}$ for S1 show good agreement with those for S2.

Figures SI-2 (a), 2(b), 2(c), and 2(d) show $T_{\text{min}}$ as a function of the tentative film thickness obtained for S1 (red circles) and S2 (green triangles) assuming that the final thicknesses $t_{final}$ are 3 ML, 4 ML, 5 ML, and 6 ML, respectively. Figures SI-3 (a), 3(b), 3(c), and 3(d) display $\rho_{5K}/\rho_{230K}$ versus the tentative film thickness obtained for S1 (red circles) and S2 (green triangles) assuming that $t_{final}$ = 3 ML, 4 ML, 5 ML, and 6 ML, respectively. These plots in Figs. SI-2 and SI-3 were drawn using the same data as shown in Figs. 2(b) and 2(c). If we assume that $t_{final}$ = 3 ML, both $T_{min}$ and $\rho_{5K}/\rho_{230K}$ as a function of the film thickness for S1 are in good agreement with those for S2, as shown in Figs. SI-2(a) and SI-3(a), respectively. However, if we assume $t_{final}$ = 4 ML, the agreement becomes slightly worse, as shown in Figs. SI-2(b) and SI-3(b). If $t_{final}$ is increased further, up to 5 ML and 6 ML, the agreement becomes even worse, as shown in Figs. SI-2(c) and SI-3(c) and Figs. SI-2(d) and SI-3(d), respectively. Thus, it is considered that the actual final thickness $t_{final}$ is 3 ML.

(a) $t_{final}$ = 3ML

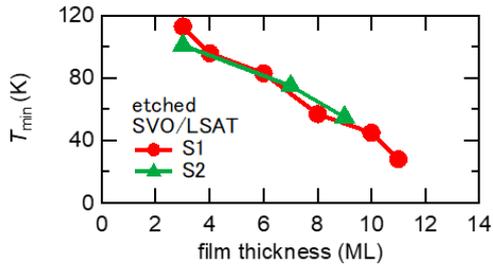

(b) $t_{final}$ = 4ML

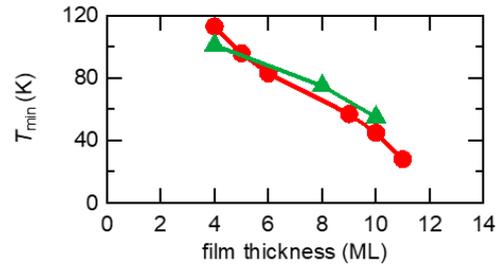

(c) $t_{final}$ = 5ML

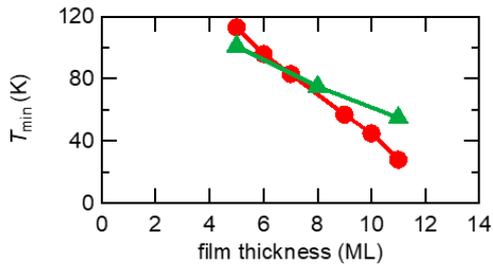

(d) $t_{final}$ = 6ML

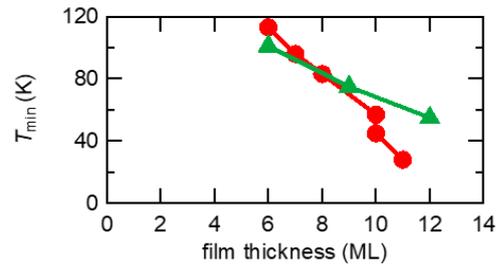

**FIG. SI-2.** $T_{min}$ as a function of the tentative film thickness for S1 (red circles) and S2 (green triangles) obtained assuming that the final thicknesses $t_{final}$ are (a) 3 ML (b) 4 ML (c) 5 ML, and (d) 6 ML. All the lines are guides to the eye.

(a) $t_{final}$ =3ML

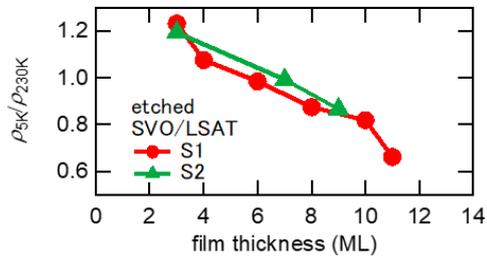

(b) $t_{final}$ = 4ML

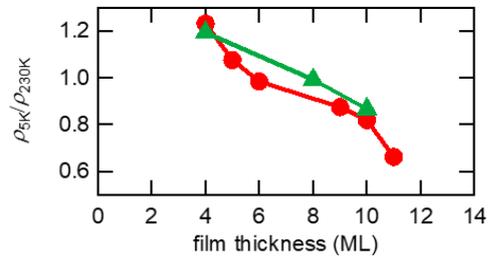

(c) $t_{final}$ = 5ML

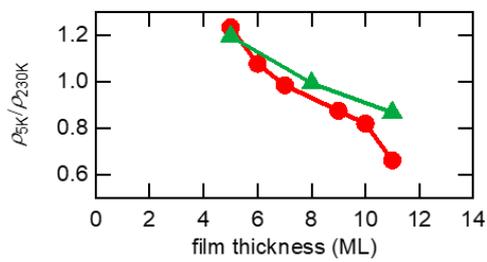

(d) $t_{final}$ = 6ML

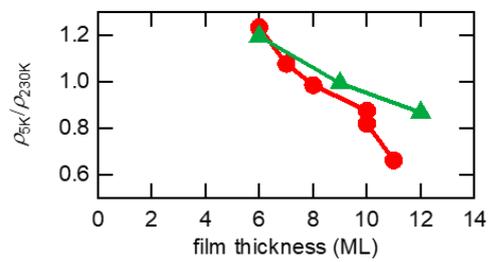

**Fig. SI-3.** $\rho_{5K}/\rho_{230K}$ as a function of the tentative film thickness for S1 (red circles) and S2 (green triangles) obtained assuming that the final thicknesses $t_{final}$ are (a) 3 ML (b) 4 ML (c) 5 ML, and (d) 6 ML. All the lines are guides to the eye.